\documentclass[aps,prl,twocolumn,showpacs]{revtex4}
\usepackage{graphicx}

\begin{document}

\newcommand{\be}{\begin{equation}}
\newcommand{\ee}{\end{equation}}
\newcommand{\e}{\epsilon}
\renewcommand{\d}{\delta}
\newcommand{\dk}{\hat{\delta}}
\newcommand{\Dk}{\hat{\Delta}}
\newcommand{\rhob}{\bar{\rho}}
\newcommand{\x}{\vec{x}}
\renewcommand{\k}{\vec{k}}
\newcommand{\g}{\gamma}
\newcommand{\df}{\hat{\delta}^{(1)}}
\newcommand{\ds}{\hat{\delta}^{(2)}}
\renewcommand{\a}{\alpha}
\renewcommand{\l}{\lambda}

\title{Limits on deviations from the inverse-square law on megaparsec scales}
\author{Carolyn Sealfon}
\email{csealfon@physics.upenn.edu}
\author{Licia Verde}
\email{lverde@physics.upenn.edu}
\author{Raul Jimenez}
\email{raulj@physics.upenn.edu}
\affiliation{
Dept. of Physics and Astronomy, University of Pennsylvania, Philadelphia,
PA 19104, USA.}

\begin{abstract}

We present an attempt to constrain deviations from the gravitational
inverse-square law on large-scale structure scales.  A perturbed law
modifies the Poisson equation, which implies a scale-dependent growth
of overdensities in the linear regime and thus modifies the power
spectrum shape. We use two large-scale structure surveys (the Sloan
Digital Sky survey and the Anglo-Australian Two-degree field galaxy
redshift survey) to constrain the parameters of two simple modifications of the
inverse-square law.  We find no evidence for deviations from normal 
gravity on the scales probed by these surveys ($\sim 10^{23}$m.)

\end{abstract}


\maketitle

\section{Introduction}

Einstein's theory of general relativity,
which generalizes Newton's law of gravity, has been extensively tested
and verified.  But precision tests of the inverse-square force law
have been performed only on scales $\le 10^{13}$ m (\cite{Adelberger}
and references therein). An enormous extrapolation is required to
apply this law to cosmological scales $> 10^{22}$m \cite{Peebles03}.

Some recent theories propose a deviation from Newtonian gravity at
large distances, perhaps due to extra dimensions, branes, or new
particles (e.g.\cite{DGP}).  These theories, mostly motivated by
finding an alternative explanation to dark energy for the cosmic
acceleration (e.g.\cite{Arkani-Hamed02,DDG02, Carroll03, Lue1}), modify
gravity on scales comparable to the horizon scale $\sim 10^{26}$m.

Here we set out to constrain theories in which gravity deviates from
the inverse-square law on scales of $\sim$10 Mpc or $10^{23}$m.
We evolve the power spectrum (and bispectrum) under a perturbed law of
gravity in the linear (and mildly non-linear) regimes, from
recombination to the present day.  The perturbed law modifies the
evolution of density fluctuations, altering, in particular, the power
spectrum shape.  We compare the prediction for the modified power
spectrum to that measured from two galaxy redshift surveys: the
Anglo-Australian Two-degree Field Galaxy Redshift Survey (2dFGRS;
\cite{Colless2dF,Percival01}) and the Sloan Digital Sky Survey (SDSS;
\cite{GunnSDSS,YorkSDSS,Teg03}).

In the models we consider, gravity is consistent with the
inverse-square law on scales smaller than $\sim$1 Mpc, so the
deviation does not affect early universe physics through
recombination.  Thus, we assume that the standard computation of the
cosmic microwave background (CMB) anisotropies holds, and that the CMB
observations provide the mass power spectrum (and bispectrum) at
recombination.  Moreover, as explained later, we marginalize over
reasonable priors for the power spectrum amplitude and the primordial
power spectrum slope, thus making our analysis insensitive to
reasonable changes in the other cosmological parameters.
We expect deviations from the inverse-square law to modify the
integrated Sachs-Wolfe (ISW) effect.  However, as this affects the CMB
on large scales where cosmic variance is large, we do not discuss ISW
here.

In the context of Newtonian gravity, a similar approach has been
widely used in the literature to constrain cosmological parameters
(e.g., \cite{Spergel03}) and biasing, the relation between clustering
of galaxies and clustering of mass (e.g.,\cite{Lahav02, Verde02}).
Here, we fix the background cosmology \cite{Spergel03} and assume a
scale-independent bias
\footnote{Justification of this assumption, used in all analyses of
large-scale structure, relies on the statistical properties of
initially gaussian random fields \cite{Kai86b, Weinberg97}.}.
The deviation from Newtonian gravity is
parameterized by a length-scale and a strength on which we place
constraints.
We find no compelling evidence for deviations from the
inverse-square force law on large-scale structure scales.
This null result leaves open the possibility that both Newton's gravity and our weak assumptions about scale-independent bias and the primordial power spectrum are incorrect in some contrived way so that their effects conspire to exactly cancel.  We regard such a scenario as exceedingly unlikely.
This is the first attempt to use cosmological data to
constrain gravity on these scales, but for related
theoretical work, see \cite{Uzan, Peebles02, Barrow98, Nusser04}.

\section{Method}
We consider a force of gravity that deviates from an inverse-square
law by a small perturbation 
on Mpc scales.
We examine two possible functional forms
for this perturbation, each parameterized by a strength and a length
scale.
Since Newtonian gravity has been tested on very
small (mm) scales up to solar system scales \cite{Adelberger}, both models recover the inverse-square law at small scales.
The expansion rate $a(t)$ is set by WMAP's best-fit $\Lambda$CDM
background cosmology \cite{Spergel03}.  
Note that these models differ from the modification to general relativity considered in \cite{Lue2}.
Since we are testing a null hypothesis, the specific form of the parameterization does not matter; our choices are based loosely on parameterizations used in small-scale tests of gravity.

In the first model, we consider a Yukawa-like contribution to the potential,
\be
\Phi(\vec{r})= -G \int d^3 r' \frac {\rho(\vec{r'})} {\vert \vec{r}-\vec{r'} \vert} \left[1 + \a\left(1-e^{-\frac{\vert \vec{r}-\vec{r}' \vert}{\l}}\right)\right],
\ee
where $\a$ is small.   The Yukawa parameterization has been used in small-scale tests of gravity, where it is physically motivated by the exchange of virtual bosons \cite{Adelberger}.  This model nicely behaves like an inverse-square law at very large and very small length scales relative to the length scale $\l$.

In the second model 
(hereafter the ``PL model''), we consider a power-law-like (PL) potential of the form, 
\be
\Phi(\vec{r})= -G \int d^3 r' \frac {\rho(\vec{r'})} {\vert
\vec{r}-\vec{r'} \vert} \left[1 + \tilde{\e}(\vert \vec{r}-\vec{r'}
\vert) \left(\frac{R}{\vert \vec{r}-\vec{r'}
\vert}\right)^{N-1}\right] \ee 
where $\tilde{\e}(r)=\Theta(r-R)$ is a step function, $\e \ll 1$, 
$R$ is a physical length scale, $\e$ is the strength of
the deviation, 
and we discuss $N=2$.  The $r$ dependence of
$\tilde{\e}$ ensures agreement with 
standard gravity at small scales.
Incidentally, this form of modification (with constant
$\tilde{\e}$) can be generated by the simultaneous exchange of two
massless scalar particles (\cite{Adelberger} and references therein).
However, such effects would likely occur at much smaller scales than
we are probing.

Let's consider the first model. In comoving coordinates, $\x \equiv \frac {\vec{r}}{a(t)}$, the
comoving gravitational potential is given by $ \phi(\vec{x})=-G \rhob
a^2 \Delta(\vec{x})$, where
\be
\Delta(\vec{x})= \int d^3 x' \frac {1+\d(\vec{x'})} {\vert \vec{x}-\vec{x'}\vert} \left[1 + \a\left(1-e^{-\frac{a \vert \vec{x}-\vec{x}' \vert}{\l}}\right)\right],
\ee
$\d(\vec{x}) = \frac {\rho(\vec{x})-\rhob}{\rhob}$, $\rhob$ is the average density of the universe, and we scale $a(t)$ so $a=1$ today.

Transforming $\Delta$ into Fourier space yields
\be
\Dk(\k) = \frac{4 \pi \dk(\k)}{k^2} \left[1 + \a F_{\Delta_1}\left(\frac{a}{k \l}\right)\right]
\ee
with $F_{\Delta_1}(s) = \frac{s^2}{1+s^2}$.

Following the same analysis as the first model, the PL model yields
\be
\label{eq:Dk}
\Dk(\k) = \frac{4 \pi \dk(\k)}{k^2} \left[1 + \e F_{\Delta_2}\left(\frac{a}{k R}\right)\right],
\ee
with $F_{\Delta_2}(s) = \frac{1}{s}\left[\frac{\pi}{2} - {\rm Si}\left(\frac{1}{s}\right)\right]$, where ${\rm Si}(x)$ denotes the sine integral.

\subsection{Linear theory}
In this section we compute the density fluctuations $\dk$, the comoving
velocity field, and the power spectrum for the Yukawa-like model. The results
for the PL model are obtained simply by replacing $\a$ with $\e$,
$\l$ with $R$, and $F_{\Delta_1}(s)$ with $F_{\Delta_2}(s)$.

The universe's expansion rate is dominated by physics at horizon
scales, beyond the scales we consider here.  We thus use Friedman
equations for the background cosmology. Although for the PL model we
simply make this assumption, for the Yukawa-like model this assumption
can be justified. The Yukawa-like model recovers the
inverse-square law on horizon scales.  A change in  the value of Newton's constant $G$, as in the large-scale limit of the Yukawa-like model, has
the effect of changing the overall amplitude of the growth factor, and
thus of the power spectrum normalization, over which we marginalize.
We are sensitive only to changes in the
shape of the power spectrum on supercluser scales, which is most
strongly affected by the growth of perturbations against the background
cosmology, and not by the background cosmology itself.

The equation for $\dk$ to first order in perturbation theory is
\be
\ddot{\dk} + 2 \frac{\dot{a}}{a} \dot{\dk} - G \rhob k^2 \hat{\Delta} = 0.
\ee
For $\a = 0$ the equation is separable, with solution\footnote{We neglect the decaying solution.}
\be
\dk_A(\k, t)\equiv A(t) \dk_o(\k)\,.
\ee

We now look for a solution of the form \be \dk(\k,t) = \dk_A(\k,t)
\left[1+\a \hat{d}(\k, t)\right] \ee to first order in $\a$, and solve
for $\hat{d}(\k, a(t))$ \footnote{The differential equation is $\hat{d}'' \!+\! \left[\!3 \frac{H'}{H}\! +\!\frac{2}{H^2 A a^3}\! +\!\frac{3}{a}\!\right]\! \hat{d}' \!+\!\left[\!\frac{H''}{H}\!+\! \left(\!\frac{H'}{H}\!\right)\!^2 \!+\!\frac{3 H'}{a H} \!-\! \frac{3}{2 a^2}\!+\!\frac{3 H_o^2 \Omega_\Lambda}{2 H^2 a^2}\!\right]\!\hat{d}\! =\!\left[\!\frac{3}{2 a^2} \!-\!\frac{3 H_o^2 \Omega_\Lambda}{2 H^2 a^2}\!\right]\!F_\Delta(\!s\!)$, where $H$ and $A$ are functions of $a$ and primes denote $\frac{d}{da}$.}.  
The initial
conditions $\hat{d}(0)=0$ and $\hat{d}'(0)=0$ ensure that gravity
behaves normally at small scales and that $\dk$ is finite at $a=0$.
For the PL model, all solutions are undefined at $a=0$.  We set
$\hat{d}$ and $\hat{d}'$ to zero at   
a sufficiently small scale that changing that scale by an order of magnitude changes $\hat{d}$ at $a=1$ by significantly less than our error bars.

\begin{figure}
\includegraphics{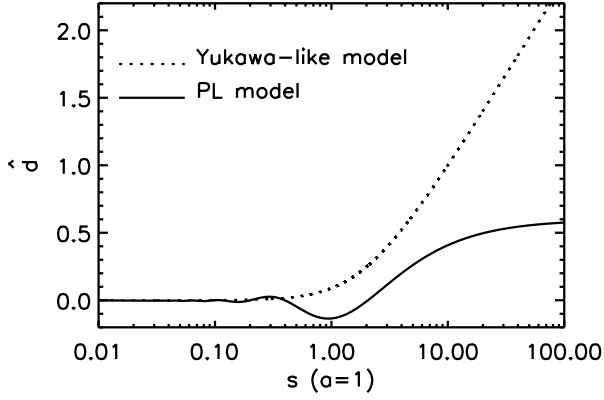}
\caption{$\hat{d}(\frac{1}{k \l})$ and $\hat{d}(\frac{1}{k R})$ for the
Yukawa-like and PL models, respectively, at $a=1$ for a $\Lambda$CDM
cosmological model ($\Omega_{m} =0.27$ and $\Omega_{\Lambda}=0.73$)}
\label{fig:dhat}
\end{figure}

Figure(\ref{fig:dhat}) shows $\hat{d}(\frac{1}{k \l})$ for the Yukawa-like
model (dashed line) and $\hat{d}(\frac{1}{k R})$ for the PL model (solid line) for a
$\Lambda$CDM universe at $a=1$.  Since a positive value of $\a$ makes
gravity stronger, we expect more clustering ($\hat{d} > 0$).
The step function in the PL model's potential causes some ringing in the Fourier transform, so the change in the comoving potential and $\hat{d}$ both oscillate as functions of $k$.

We find that  $\Omega_{\Lambda}$ weakens the effect of $\a$ or $\e$ at large scales.

We can now compute the comoving velocity to first order in
perturbation theory using the continuity equation and the equation of
motion in Fourier space \cite{Peebles}. For simplicity, we'll work in
an Einstein-de Sitter (EdS) universe.  We only use the velocity to compute
the bispectrum, and the bispectrum is robust to changes in cosmology
(see, e.g., Figure (1) in \cite{MVH97}).
 To first order in $\a$, we
find

\be
\hat{\vec{v}} = i a \frac{\k}{k^2} \dot{\dk} = i a \frac{\k}{k^2} \dot{A} \dk_o \left[ 1+ \a F_v(\frac{a}{k \l})\right],
\ee
where
\be
F_v(s) \equiv \hat{d}(s) + s \hat{d}'(s).
\ee

The power spectrum is given by 
\be
(2 \pi)^3 P(\k,t) \delta^D(\k +\k')=\langle \dk(\k,t) \dk(\k',t) \rangle,
\ee
where
\be  
\langle \vert \dk(\k, t) \vert^2 \rangle  = \langle \vert \dk_A(\k,t) (1 + \a \hat{d}(\k, t))\vert^2 \rangle
\ee
and $\langle \rangle$ denote an ensemble average over all realizations of the universe. 
The present-day power spectrum derived using our model of modified gravity,
to first order in $\a$, is then
\be
\label{eq:P}
P(k)= P_o(k) \left[ 1 + 2 \a \hat{d}\left(\frac{1}{k \l}\right)\right],
\ee
where $P_o$ is the power spectrum expected for Newtonian gravity.

To test the validity of our first-order approximation in $\a$, we also compared $\hat{d}$ for an EdS universe to the
Yukawa-like model's exact solution.  The growing EdS solution for the density fluctuation is a hypergeometric function,
\be
\dk(s)=
_2\!F_1\!\left(\!\frac{5-\sqrt{25 \!+\! 24\a}}{8},\frac{5+\sqrt{25 \!+\! 24\a}}{8}, \frac{9}{4};\!-s^2\!\right) s\, \dk_0(\!k\!),
\ee
where $s \equiv \frac{a}{k \l}$.  We find that the higher-order $\a$ corrections contribute a small
fraction most values of $s=\frac{a}{k\l}$ in our analysis.  For
example, for $s=100$ (the largest $s$ in our analysis), we get $\sim 20$\% error for
$|\a| \sim 0.2$; for small values of $s$ --and the same $|\a|$ value-- the error is less than a percent.
We are only concerned with effects that change the result by orders of
magnitude, not factors of a few, since gravity has never been tested
on these scales before.  Thus, we believe the Taylor expansion in $\a$
and $\e$ is sufficient for our first search for constraints.

\begin{figure*}[ht]
\includegraphics{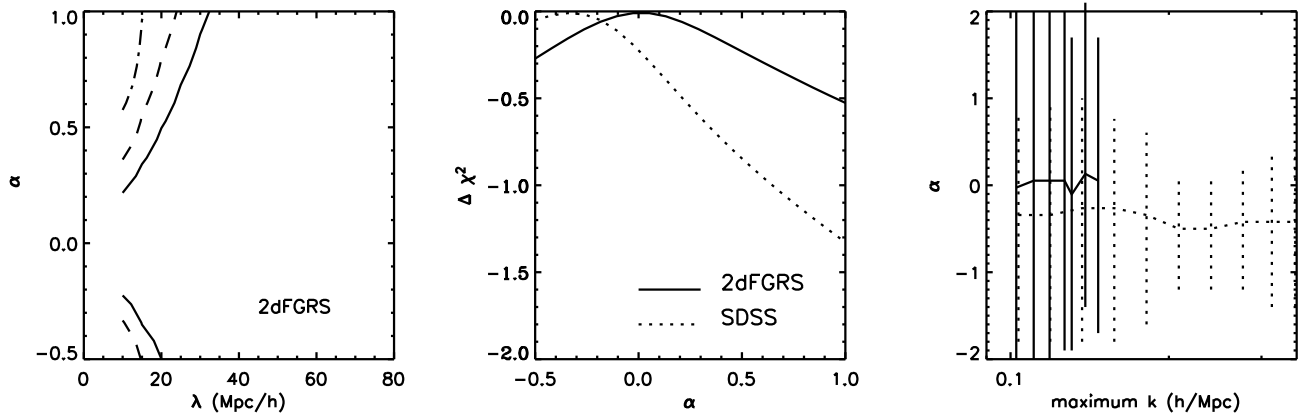}
\caption{Results from the first (Yukawa-like) model. Left:
Likelihood as a function of $\a$ and $\l$ using data up to $k \sim
0.15$ h/Mpc using 2dF data.  Contours denote one sigma marginalized 
(solid line), one sigma joint (dashed line),
and two sigma joint (dashed-dotted line). Middle: Chi square as function of $\a$,
marginalized over $\l$, using data up to $k \sim 0.15$. Right: Values
of $\a$ corresponding to the maximum chi square with error bars
denoting one sigma, as a function of the maximum $k$ included in the
analysis.}
%  Solid lines denote 2dF results and dashed lines denote SDSS results.}
\label{fig:yukawa}
\end{figure*}

\subsection{Mildly non-linear theory}
The equation for $\dk$ to second order in perturbation theory is

\be
\ddot{\dk} + 2 \frac{\dot{a}}{a} \dot{\dk} + \frac{k^2}{a^2} \hat{\phi} = -\frac{1}{a^2} [ k_i \hat{\phi} * k^i \dk + \dk * k^2 \hat{\phi} + k_i k_j (\hat{v}^i * \hat{v}^j)],
\ee
where we use the Einstein summation convention over spacial indices $i,j$.  The symbol ``$*$'' denotes a convolution in $k$-space,

\be
\label{eq:convolution}
f(\k)*g(\k) \equiv \int d^3l \int d^3m \, \delta^D (\vec{l}+\vec{m}-\vec{k}) f(\vec{l}) g(\vec{m}),
\ee
where $\delta^D(\k)$ is the three-dimensional Dirac delta function.  We use the solution for $\vec{v}$ from first-order perturbation theory.

Let $\dk^{(1)}$ represent the solution to $\dk$ in linear theory, and $\dk^{(2)}$ represent the second-order term (of order ${\dk_A}^{\,2}$), so $\dk = \dk^{(1)} + \dk^{(2)}$ plus smaller, higher-order terms.

For $\a=0$ (or $\e=0$), the solution for the second-order term is
\begin{eqnarray}
\dk_A^{(2)} &\equiv& A^2(t) \dk_o^{(2)}(\k)
\\
&=& A^2\!\left(\! \frac{5}{7} \dk_o * \dk_o + k_i \dk_o * \frac{k^i}{k^2} \dk_o + \frac{2}{7} \frac{k_i k_j}{k^2} \dk_o * \frac{k^i k^j}{k^2} \dk_o\!\right). \nonumber
\end{eqnarray}

We then follow the same approach as in the linear theory.  Using the first model as our example, we look for a solution to first order in $\a$ of the form
\be
\dk^{(2)} = \dk_A^{(2)}[1 + \a \hat{g}(\k,t)],
\ee
Switching the time variable to $a$, the resulting differential equation for $\hat{g}(\k, a)$ in an EdS universe is
\be
\label{eq:diffg}
\frac{\partial^2 \hat{g}}{\partial a^2}+\frac{11}{2 a} \frac{\partial \hat{g}}{\partial a} + \frac{7}{2 a^2}\hat{g}=\frac{1}{\dk_o^{(2)}} \frac{3}{2 a^2} N(\k, a),
\ee
where
\begin{eqnarray}
\label{eq:N}
N(\k,a)&=&F_{\Delta_1}\left( \frac{a}{k\l}\right) \dk_o^{\left(2\right)}
\nonumber\\
&+& \frac{k_i}{k^2} \dk_o \!\left(\!F_{\Delta_1}\!\left(\! \frac{a}{k \l}\!\right)\! + \!\hat{d}\!\left(\! \frac{a}{k \l}\right) \!+\! \frac{4}{3} F_v\!\left(\! \frac{a}{k \l}\!\right)\!\right)\!*k^i \dk_o 
\nonumber\\
&+& \frac{k_i}{k^2} \dk_o * k^i \dk_o \left(\hat{d}\!\left( \frac{a}{k \l}\right)+ \frac{4}{3} F_v\!\left( \frac{a}{k \l}\right)\right) 
\nonumber\\
&+& \dk_o*\dk_o \!\left(\!F_{\Delta_1}\!\left(\! \frac{a}{k \l}\!\right)\!+ \!2 \hat{d}\!\left(\! \frac{a}{k \l}\!\right)\! +\!\frac{4}{3}F_v\!\left(\! \frac{a}{k \l}\!\right)\!\right)
\nonumber\\
&+& \frac{4}{3} \frac{k_i k_j}{k^2} \dk_o F_v\!\left( \frac{a}{k \l}\right)* \frac{k^i k^j}{k^2} \dk_o,
\end{eqnarray}
with initial conditions  $\hat{g}(\k, 0)=0$ and $\frac{\partial}{\partial a} \hat{g}(\k, 0)=0$.
The bispectrum is given by
\begin{eqnarray}
B(\k_1,\k_2)& \delta^D(\k_1+\k_2+\k_3) \equiv \langle \dk(\k_1) \dk(\k_2) \dk(\k_3)\rangle
\nonumber\\
&= \langle [\df(\k_1)+\ds(\k_2)][\df(\k_2)+\ds(\k_2)]
\nonumber\\
&\times [\df(\k_3)+\ds(\k_3)]\rangle.
\end{eqnarray}   
The leading term, $\langle \df(\k_1) \df(\k_2) \df(\k_3) \rangle$,
vanishes under Gaussian initial conditions.  Modified gravity still preserves gaussianity in the linear regime, and introduces a correction to the bispectrum in second-order perturbation theory.  To first non-vanishing
order in $\dk_A$ and $\a$, we get
\begin{eqnarray}
\langle \dk(\k_1) \dk(\k_2) \dk(\k_3)\rangle=A^4 \langle \dk_o(\k_1)\dk_o(\k_2) \ds_o(\k_3) \rangle 
\nonumber\\
\!\!\times\! \left[\!1\!+\!
\a \!\left(\!\hat{d}(\frac{a}{k_1 \l})+\hat{d}(\frac{a}{k_2 \l})+\hat{g}(\k_3,a)\right)\!\right]\!+\!cyc.
\end{eqnarray}
where again we can pull $\hat{d}$ and $\hat{g}$ out of the ensemble average, $\k_3=-\k_1-\k_2$, and $cyc.$ denotes cyclic permutations of the subscripts 1,2,3.

\begin{figure*}[ht]
\includegraphics{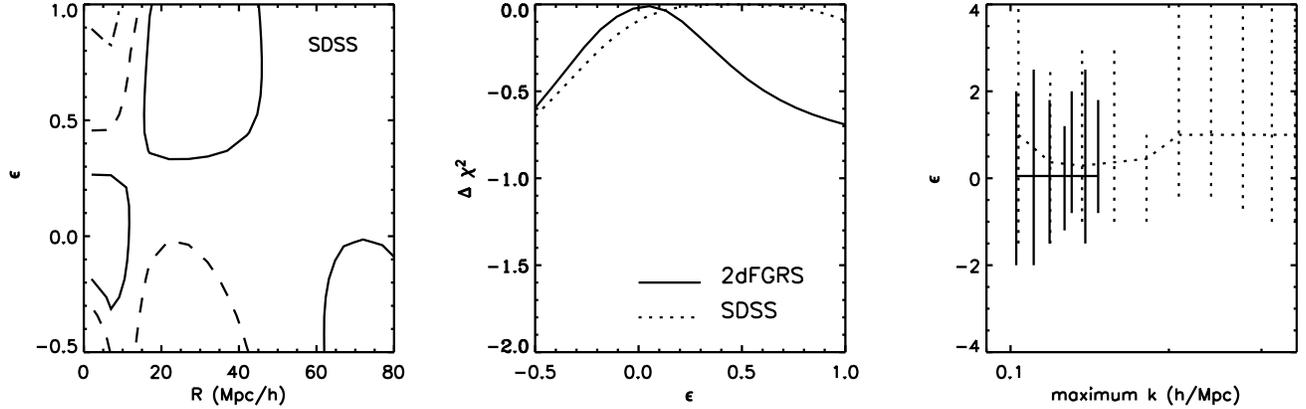}
\caption{Results from the PL model. Left:
Likelihood as a function of $\e$ and $R$ using data up to $k \sim
0.15$ h/Mpc using SDSS data.  Inner contours denote one sigma marginalized 
($\sim \Delta \chi^2= 1$ solid line), bottom contour denotes one sigma
joint ($\sim \Delta \chi^2=2.3$ dashed line). Right: Chi square as function of $\e$,
marginalized over $R$, using data up to $k \sim 0.15$. }
\label{fig:N2}
\end{figure*}

Since we already know $\hat{d}$, we need to solve only for the last term.  To do so, let's examine the term  $X \equiv \langle \dk_o(\k_1)\dk_o(\k_2) \ds_o(\k_3) \hat{g}(\k_3,a) \rangle$. From Eq.(\ref{eq:diffg}) we can write,
\be  
\label{eq:diffX}
\frac{\partial^2 X}{\partial a^2}+\frac{11}{2 a} \frac{\partial X}{\partial a} + \frac{7}{2 a^2}X=\langle \dk_o(\k_1)\dk_o(\k_2) \frac{3}{2 a^2} N(\k_3, a) \rangle.
\ee
$N(\k,a)$ is a sum of terms with k-space convolutions and $a$ dependence.  We can thus rewrite  Eq.(\ref{eq:N}) as
\be
\label{eq:N2}
N(\k,a) = \sum_n \dk_o(\k) J_n(\k,a) * \dk_o(\k)K_n(\k,a),
\ee
where the expressions for $J_n$ and $K_n$ are easily obtained by equating Eq.(\ref{eq:N2}) with Eq.(\ref{eq:N}).
Using Eq.(\ref{eq:convolution}), we rewrite the right hand side of Eq.(\ref{eq:diffX}) as
\begin{eqnarray}
\frac{3}{2 a^2} &\langle \int dl^3 dm^3 \delta^D(\vec{l}+\vec{m}-\k_3) \dk_o(\k_1) \dk_o(\k_2) \dk_o(\vec{l}) \dk_o(\vec{m})\nonumber\\
&\times \sum_n J_n(\vec{l},a) K_n(\vec{m},a) \rangle
\end{eqnarray}
which becomes
\be
\frac{3}{2 a^2} \frac{P_o}{A^2}(\!k_1\!)\frac{P_o}{A^2}(\!k_2\!)\sum_n \! J_n(\k_1,a) K_n(\k_2,a) \!+\! J_n(\k_2,a) K_n(\k_1,a). \nonumber
\ee
We can then solve Eq.(\ref{eq:diffX}) for each function of $a$ that appears in the above sum, and get an expression for the bispectrum. 

The full expression for the bispectrum given by,
\be
B(\k_1, \k_2)\delta^D\!(\k_1+\k_2+\k_3)=J(\k_1,\k_2)P_o(k_1)P_o(k_2) +cyc.\,,
\ee
where
\begin{eqnarray}
J(\k_1,\k_2)&=&J_o(\k_1,\k_2)+\a
\left(J_o(\!\k_1,\k_2\!)\right.\!\left(\!\hat{d}\!\left(\!\frac{a}{k_1
\l}\!\right)+\hat{d}\!\left(\!\frac{a}{k_2
\l}\!\right)\!\right)\nonumber\\ 
&+&\left[G_1\!\left(\!\frac{a}{k_3
\l}\!\right)\left(\frac{5}{7} +
\frac{\k_1\cdot\k_2}{k_2^2}+\frac{2}{7}\left(\frac{\k_1 \cdot
\k_2}{k_1 k_2}\right)^{\!2}\right) \right.\nonumber\\ 
&+&\frac{\k_1 \cdot
\k_2}{k_1^2} \left(G_1\!\left(\!\frac{a}{k_1 \l}\!\right) +
G_2\!\left(\!\frac{a}{k_1 \l}\!\right)+\frac{4}{3}G_3\!\left(\!\frac{a}{k_1
\l}\!\right) \right.\nonumber\\ &&+ \left. G_2\!\left(\frac{a}{k_2
\l}\right) +\frac{4}{3}G_3\!\left(\frac{a}{k_2 \l}\right)\right)
\nonumber\\ &+& G_1\!\left(\frac{a}{k_2\l}\right)+ 2
G_2\!\left(\frac{a}{k_2 \l}\right)+ \frac{4}{3}G_3\!\left(\frac{a}{k_2
\l}\right) \nonumber\\ 
&+&
\left. \left. \frac{4}{3}G_3\!\left(\!\frac{a}{k_2
\l}\!\right)\left(\!\frac{ \k_1 \cdot \k_2}{k_1 k_2}\!\right)^{\!2} \!+ (\k_1
\!\leftrightarrow\! \k_2) \!\right] \right),
\end{eqnarray}
$J_o(\k_1,\k_2)P_o(k_1)P_o(k_2)$ is the Newtonian gravity result given by Eq.(40) of \cite{Fry84}, and
\begin{eqnarray}
G_1(s)&=& \frac{3}{7} +\frac{2}{5 s^2} +\frac{3 \sqrt{2}}{10} \frac{1}{s^{\frac{7}{2}}}
\arctan\left(\frac{\sqrt{2 s}}{s - 1}\right)  \nonumber\\
&-& \frac{3}{5 s}\arctan(s) +\frac{3\sqrt{2}}{20} \frac{1}{s^{\frac{7}{2}}} \ln\left(\frac{1+s+\sqrt{2 s}}{1+s-\sqrt{2 s}}\right), \nonumber\\
G_2(s)&=&\frac{9}{50 s^{\frac{7}{2}}} \int_0^s \!\!dx\,x^\frac{5}{2}\!\!\left(_2F_1^{(1,0,0,0)}\!\!\left(\!0, \frac{5}{4},\frac{9}{4}; -x^2\!\right) \right.\nonumber\\ 
&&-\left. _2F_1^{(0,1,0,0)}\!\!\left(\!0, \frac{5}{4},\frac{9}{4}; -x^2\!\right)
\right) \nonumber\\
&-& \frac{9}{50 s} \int_0^s \!dx \left(
_2F_1^{(1,0,0,0)}\!\!\left(\!0, \frac{5}{4},\frac{9}{4}; -x^2\!\right) \right. \nonumber\\
&&- \left. _2F_1^{(0,1,0,0)}\!\left(\!0, \frac{5}{4},\frac{9}{4}; -x^2\!\right)
\right) , \nonumber\\
G_3(s)&=&\frac{1}{s} \!\int_0^s \!\!dx \frac{x^2}{5} \,\!_2F_1\!\left(1, \frac{9}{4},
\frac{13}{4}; -x^2\right) %\right.\nonumber\\
\nonumber\\
&+&\frac{1}{s^{\frac{7}{2}}} \!\int_0^s \!\!\!dx \frac{-x^\frac{9}{2}}{5} \,\!_2F_1\!\left(1,
\frac{9}{4}, \frac{13}{4}; -x^2\right) %\right. \nonumber\\
\nonumber\\&+& G_2(s),
\end{eqnarray}
Here, $_2F_1^{(1,0,0,0)}$ denotes the partial derivative of the hypergeometric function with respect to the first argument, and
$_2F_1^{(0,1,0,0)}$ denotes the partial derivative with respect to the second argument.

 Since the bispectrum contains complementary information from the
 power spectrum, we anticipate that combining the two should produce
 stronger constraints. However, we leave the bispectrum analysis to
 future work.

\section{Analysis}
We can use the power spectrum measurements from the 2dFGRS \cite{Percival01} and the  SDSS \cite{Teg03} to
constrain the parameters of our models.  
The bispectrum derivation shows that the bispectrum should help disentangle the effects of biasing from possible modifications to gravity, but the analysis will follow in a subsequent paper.

The $z=0$ power spectrum for normal gravity can be theoretically
calculated using the transfer functions computed with CMBFAST for the
WMAP best-fit parameters.  The theoretical
power spectrum for given values of $\a$
and $\l$ (or $\e$ and $R$) can then be obtained from Eq.(\ref{eq:P}).
Assuming Gaussian likelihood, we computed the likelihood for each
survey's power spectrum measurements as a function of $\a$ and $\l$
(or $\e$ and $R$).

To make our analysis insensitive to assumptions about the background cosmology, we marginalized over reasonable priors for
the power spectrum amplitude and the primordial power-law power
spectrum slope, $n$.  For the amplitude, we find that the ranges 0.6 to 1.4 (for the
first model) and 0.4 to 1.4 (for the PL model) times the best-fit power 
spectrum amplitude for normal gravity are sufficient 
to allow for changes in the expansion rate of the universe.
The values of $\Omega_m$, $\Omega_{\Lambda}$, and the dark energy
equation of state parameter (assumed to be constant) affect primarily
the power spectrum amplitude in linear theory and for the scales of
interest. 
We marginalized over $n$ using the WMAP prior ($n=0.99\pm 0.04$ at the
one-sigma level \cite{Spergel03}).
The large scale structure power spectrum shape has a weak dependence on $\Gamma\simeq
\Omega_m h$, which shows up only on the largest scales and is strongly
degenerate with $\Omega_{baryon}$ and $n$ \cite{Percival01,Pope04}.
The large-scale structure power spectrum has a very weak sensitivity
to $\Omega_{baryon}$ via two effects.  One is the appearance of the
so-called baryonic wiggles; as this effect has not yet been detected
from these data sets, it cannot yet give sensitivity to
$\Omega_{baryon}$. The other effect is a weak change in power spectrum
slope, degenerate with $n$.
We are also somewhat insensitive to the uncertainty in neutrino mass.
Increasing neutrino mass affects first the amplitude and then the
shape of the galaxy power spectrum in the scales of interest (see,
e.g., figure(8) of \cite{Verde03}).
Present constraints would be weakened if a running of
$P(k)$ and/or a redshift dependence of the dark-energy equation of
state were explored.

Lastly, since we are mainly interested in a
constraint on $\a$ and/or $\e$ rather than on the scale $\l$ and/or
$R$, we marginalized over $\l$ (or $R$) from 10 to 100 Mpc/h.

\section{Results and Conclusions}
Figure (\ref{fig:yukawa}) shows the Yukawa-like model's parameter
constraints for the 2dFGRS and SDSS data, and figure (\ref{fig:N2})
shows the PL model's constraints.  The ``ringing'' in the $\hat d$
function in the PL model, creates ``wiggles'' in the large scale
structure power spectrum which are responsible for the geometry of the
iso-likelihood contours in fig 3.  The left plots show the
likelihood contours for one-sigma marginalized, one-sigma joint, and
two-sigma joint confidence levels
 for one survey in $\l$,$\a$ (or $R$,$\e$) space.
The center plots show the difference in chi square ($\Delta \chi^2$)
as a function of $\a$ or $\e$, marginalized over $\l$ or $R$.  We used
data up to $k \sim 0.15$ h/Mpc for both surveys.  The results are
consistent with each other and with a strict inverse-square law.
Since the constraints are weak and the Taylor expansion analysis breaks down at large values
of $|\a|$ and $|\e|$, we extrapolated the error bars from
the graphs.  
We find
$\a =0.025 ^{+1.7}_{-1.7}$, $\e = 0.05 ^{+1.3}_{-0.4}$ for 2dFGRS 
and $\a = -0.35^{+0.9}_{-0.9}$, $\e = 0.45 ^{+1.4}_{-1.4}$ for SDSS, at the one-sigma level.  Non-linearities become increasingly important as $k$
increases and might introduce systematic errors, but the signal-to-noise for the power spectrum increases with increasing $k$.  Therefore we
explore the dependence of the result as a function of the maximum $k$
included in the analysis ($k_{\rm max}$) on the right panel of the figures.

Since our analysis probes deviations in the {\it shape} of today's
power spectrum, the strongest constraints on $\a$ and $\e$ occur at
scales where the {\it slope} of $\hat{d}$ (with $a=1$) is greatest.
This typically corresponds to $\lambda$ and $R$ towards the smallest
scales of the data, so the smallest comoving scales probed have just
reached the physical scale of the force-law deviation today, and the
larger comoving scales have passed it.  (A deviation near a physical
scale $\lambda$ or $R$ has only affected comoving scales that have
grown larger than $\lambda$ or $R$.)
The weakness of our constraints arises primarily from the need to marginalize over a wide range of power spectrum amplitudes and the spectral slope.
Future weak lensing surveys will measure the power spectrum for dark
matter directly.  If $a(t)$ can also be measured independently, this
data will eliminate the need to marginalize over the amplitude and
shrink our error bars for the PL model.  We forecast that this would
roughly halve the error on $\e$.  The constraint does not improve significantly
for the Yukawa-like model.

\begin{acknowledgments}
We thank David Spergel for discussions that stimulated this 
investigation, and  Sean Carroll, Mark Trodden, Alan Heavens, and Bhuvnesh Jain for 
comments that helped improve the manuscript.
This research was supported in part by NSF0206231.
\end{acknowledgments}

\end{document}